\documentclass[12pt]{article}

\def\blackfonts{
 \font\blackboard=msbm10 scaled\magstep1
 \font\blackboards=msbm8
 \font\blackboardss=msbm6
}

\def\yblack{
  \blackfonts
  \newfam\black
  \textfont\black=\blackboard
  \scriptfont\black=\blackboards
  \scriptscriptfont\black=\blackboardss
  \def\ZZ{{\fam\black\relax Z}}
  \def\NN{{\fam\black\relax N}}
  \def\CC{{\fam\black\relax C}}
  \def\RR{{\fam\black\relax R}}
  \def\QQ{{\fam\black\relax Q}}
  \def\PP{{\fam\black\relax P}}
}
\yblack

\begin{document}

\def\ket#1{|\,#1\,\rangle}
\def\bra#1{\langle\, #1\,|}
\def\braket#1#2{\langle\, #1\,|\,#2\,\rangle}
\def\proj#1#2{\ket{#1}\bra{#2}}
\def\expect#1{\langle\, #1\, \rangle}
\def\trialexpect#1{\expect#1_{\rm trial}}

\def\openone{\leavevmode\hbox{\small1\kern-3.8pt\normalsize1}}

\title{From Quantum State Targeting to Bell Inequalities}
\author{
H. Bechmann-Pasquinucci
\\
\small
{\it {\rm UCCI.IT}, via Olmo 26, I-23888 Rovagnate, Italy}}
\date{25 February, 2004}
\maketitle

\vspace{0.5cm}
\begin{center}
{\bf Dedicated to Asher Peres on his 70th birthday}
\end{center}

\abstract{Quantum state targeting is a quantum game which results from 
combining traditional quantum state estimation with additional classical 
information. We consider a particular version of the game and show how 
it can be played with maximally entangled states. The 
optimal solution of the game is used to derive a Bell inequality for two 
entangled qutrits. We argue that the nice properties of the inequality 
are direct consequences of the method of construction.} 
\vspace{1 cm} 
\normalsize

\section {Introduction} 
Many areas of quantum mechanics have received renewed interest in the 
framework of quantum information theory. Here we will consider two 
aspects: quantum state estimation \cite{Helstrom,peresbook} and Bell 
inequalities \cite{bell,peres1}. 
These areas 
have been given completely new meaning in the light of quantum 
information theory.

Traditionally quantum state estimation is considered in the following 
way: A system is prepared in one of a known set of states and is given 
to an estimator. The task of the estimator is to identify as best as 
possible the 
state of the system and make an announcement about the state. Already 
here there are choices: the estimator could decide that he wants to 
make the best guess on the identity of every system submitted to him, in 
which case he will optimize his procedure to give him the maximum 
probability for guessing correctly the state --- and sometimes making an 
error \cite{Helstrom,peresbook}. But it could also be that it is not 
acceptable to make any 
errors, 
in which case the estimator can optimize his procedure such that when he 
identifies the state it is with certainty, by paying the price of 
sometimes obtaining an inconclusive answer \cite{peresbook,peres2}. 

With the development of quantum information theory, 
quantum state estimation has in many cases been 
given a twist. It is no longer simply a question of identifying a 
particular state out of a set of states, but there might be additional 
classical information available after the interaction with the 
'unknown' quantum state or even after a measurement has actually been 
performed. For example, in the BB84 protocol \cite{bb84} for quantum 
cryptography \cite{6state1}-\cite{dd} the eavesdropper knows that the 
quantum 
system is 
prepared with equal probability in a state belonging to a set of states 
made by two mutually unbiased bases. But she also knows that after her 
eavesdropping, i.e. 
after the interaction with 
the 'unknown' quantum state, she will  
learn in which of the two bases the system was originally 
prepared. This means that the eavesdropper knows that she will later 
receive additional classical information, which she can use to gain more 
information about the initial state of the system. 

This example with quantum state estimation in connection with quantum 
cryptography, also shows how quantum state estimation in the field of 
quantum information often becomes only a part of a bigger picture. And 
naturally this also means that different aspects are added to the 
subject, for example in eavesdropping in quantum cryptography it is no 
longer just a question of making the best identification of the 
'unknown' quantum system --- it should also be done causing minimum 
disturbance! This is the problem which lies at the center of quantum 
cryptography, namely that any interaction with the 'unknown' state which 
will lead to a higher probability of identifying it correctly will 
automatically lead to a disturbance of the state. This is what 
makes 
quantum cryptography safe for the legitimate users, since any 
eavesdropping attempt can be detected.  

One can imagine different ways to combine the elements of quantum state 
estimation and additional classical information: recently it was 
presented in a new form called 'quantum state targeting' \cite{RS}. 
Briefly described, quantum state targeting works 
as follows: There exist a set of target states which is known  to both 
players, called Alice and Bob. The task of Alice is to prepare a quantum 
system and submit it to Bob, after that Bob will reveal the
target state 
he has chosen. After receiving this information Alice makes an 
announcement concerning the 
system she prepared and finally Bob performs a fail/pass test on the 
quantum system in accordance with Alice's statement. 

Clearly, Alice uses her complete knowledge of the set of target states 
to prepare the quantum system she sends to Bob.
However, there are many strategies she could adopt to prepare the 
system and  her choice depends on what she really 
wants to
achieve. She may just optimize her control, which means the probability
that her quantum system will pass the test for being the target state
chosen by Bob. She could also be interested in optimizing the
control-disturbance trade-off. And there is also the possibility that she
is given the option to decline to make an announcement. 

It is immediately clear that there can be many different setups and
situations both involving pure and mixed states, several of which have
already received attention \cite{RS}. Here we are only interested in the
particular situation where the set of target states corresponds to two
non-orthogonal pure states, Alice goes for maximum control but at the
same time she has the option to decline to make an announcement.

It is possible to view quantum state targeting as a game by itself, but
it can also be incorporated into other games like for example weak coin
flipping \cite{RS} or it can be related with different parts of quantum
mechanics.  Recently there has been a big interest in investigating Bell
inequalities in higher dimensions \cite{mybell1,mybell2},
\cite{otherbell1}-\cite{otherbell4}.  Here we will show how playing the
game of quantum state targeting using maximally entangled states can 
lead to one of the generalized Bell
inequality for qutrits which was recently presented
\cite{mybell1,mybell2}.

Bell inequalities is indeed another area which has received renewed 
attention in the last years. Bell inequalities used to belong to the 
discussion of the completeness of quantum theory, and entanglement was 
viewed as a puzzle or even a problem to try to get rid off and not as a 
fantastic resource waiting to be explored \cite{redbook}. But that 
changed dramatically 
with the birth of quantum information theory. Suddenly entanglement had 
to be explored, characterized and application had to be 
discovered, and 
in this process Bell inequalities too, became useful tools for example 
as a security measure in quantum cryptography 
\cite{bellsec1,mybell2,bellsec2} or 
identification of 
useful correlations between quantum systems \cite{use}.

The paper is organized as follows; in section two quantum state 
targeting is defined, and a special situation is investigated. In sec. 3 
we show how to play the game of quantum state targeting by using 
maximally entangled states. In sec. 4 we establish a measure of how 
well 
Alice is playing the game of quantum state targeting and 
show that what we have obtained is actually a Bell inequality. Section 
5 is devoted to a discussion of the obtained inequality and we present 
some intuitive arguments about it's optimality. Section 6 is left for 
concluding remarks.

\section{Quantum state targeting}
The concept of quantum state targeting was recently introduced 
\cite{RS}, it is a particular way of combining quantum state 
estimation 
with additional classical information. In this section we review the basic 
idea of 
quantum state targeting and the results which are needed later.
The rules are 
quite simple and 
they are here given as they were originally defined: First 
Alice and Bob decide on a set of target states, which means that the set 
of target states is known to both of them. After this initial setup, the 
game goes in the following steps:

$~$

\noindent
(1) Alice submits a system to Bob\\
(2) Alice learn the identity of the target state (Bob reveal which of 
the possible target states he has chosen)\\
(3) Alice announces a state to Bob (a state from the set of target 
states, but not necessarily the target state chosen by Bob)\\
(4) Bob performs a pass/fail test for the state announced by Alice

$~$

\noindent
the possible outcomes of this game are the following:

$~$

\noindent
(A) Alice announces the target state and passes Bob's test\\
(B) Alice announces the target state and fails Bob's test\\
(C) Alice announces a non-target state and passes Bob's test\\
(D) Alice announces a non-target state and fails Bob's test\\

Here we are interested in a very simple situation, namely where the set 
of target states corresponds to two pure non-orthogonal states.
In this case the target state 
is chosen uniformly from a pair of two (non-orthogonal) pure states, 
denoted $\ket{{\psi}_{1}}$ and $\ket{{\psi}_{2}}$. Which means that the 
situation is the following; Alice prepares a quantum state and sends it 
to 
Bob, who will then tell Alice if he has chosen target state 
$\ket{{\psi}_{1}}$ or $\ket{{\psi}_{2}}$. Alice will then make an 
announcement about the system that she prepared and Bob will finally 
perform a pass/fail test according to Alice's statement.

Alice, of course, knows that Bob is going to chose either
$\ket{{\psi}_{1}}$ or $\ket{{\psi}_{2}}$ as target state, which means that
she can take this information into account when she prepares her quantum
state. There are several ways of doing this, depending on what Alice wants
to achieve. The situation of interest here, is where Alice goes for
maximum control, which means that she wants to optimize the probability
that Bob tests and actually finds the target state. This corresponds to
situation (A)-(B): Alice always announces the target state.  

In this situation what Alice has to do is to prepare her system
in such a way, that the probability that the system will pass the test 
for
the target state is maximal and independent of which target state Bob
chose. In \cite{RS} this particular situation was considered. It turns 
out 
that in order for Alice to have maximum control she has to submit a 
state which surprisingly enough is a pure state. Moreover, this pure 
state corresponds to what in 
connection with quantum cryptography \cite{bb84}-\cite{dd} and Bell 
inequalities 
\cite{mybell1,mybell2} usually is denoted the intermediate state 
\cite{is}. 
The intermediate state, $\ket{{\psi}_{int}}$, is precisely what the name 
indicates, namely 
the 
state which lies exactly between the two states $\ket{{\psi}_{1}}$ and 
$\ket{{\psi}_{2}}$, it can be defined as follows:
\begin{eqnarray}
\ket{{\psi}_{int}}=\frac{1}{\sqrt{\cal N}}(\ket{{\psi}_{1}} +
{e}^{-i{\phi}}\ket{{\psi}_{2}})
\end{eqnarray}
where ${\cal N}=2(1+|\braket{{\psi}_{1}}{{\psi}_{2}}|)$ is the 
normalization constant and the phase comes from 
the 
overlap 
$\braket{{\psi}_{1}}{{\psi}_{2}}= 
{e}^{i\phi}|\braket{{\psi}_{1}}{{\psi}_{2}}|$, 
and assures that when computing the 
overlap between the the states $\ket{{\psi}_{1}}$, 
$\ket{{\psi}_{2}}$ and the intermediate state both terms have the same 
phase. 

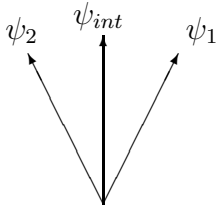
\begin{figure}[t]
\begin{center}
\begin{picture}(50,30)(0,0)
\put(25,0){\vector(1,2){10}}
\put(25,0){\vector(-1,2){10}}
\put(25,0){\vector(0,1){22.36}}
\put(12,22){${\psi}_{2}$}
\put(36,22){${\psi}_{1}$}
\put(21,24){${\psi}_{int}$}
\end{picture}
\end{center}
\caption{The states ${\psi}_{1}$, ${\psi}_{2}$ and their 
intermediate state ${\psi}_{int}$}
\end{figure}

By submitting the intermediate state Alice's control is maximized and  
 identical for the two target states and it is given by
\begin{eqnarray}
C_{max}= P({\psi}_{i}|{\psi}_{int}) 
=\frac{1}{2}(1+|\braket{{\psi}_{1}}{{\psi}_{2}}|)~~~~~~{\rm for}~~i=1,2
\end{eqnarray}
which naturally corresponds to the probability that Bob will find the
target state. 

Notice that this result is independent of the dimension of the quantum 
system.

\section{Quantum State Targeting with maximally entangled states}
Quantum state targeting as defined in the previous section can either be 
considered a game by itself or it can be incorporated into other 
games, for example weak coin flipping \cite{RS}. Here we are interested 
in 
making the connection between quantum state targeting and Bell 
inequalities, in order to show that a Bell inequality at least in some 
cases can be derived as the result of playing a quantum game. This means 
that the next step is to introduce the concept of entanglement in 
connection with quantum state targeting, since until now the game has 
been played on single systems. In this section we show how the results 
from the previous section can be adopted in the case where Alice and 
Bob share maximally entangled states.

Suppose that Alice and Bob share many maximally entangled pairs of 
qutrits,
\begin{eqnarray}
\ket{\psi_t}&=&\frac{1}{\sqrt{3}}(\ket{a_0,a_0}+\ket{a_1,a_1}+\ket{a_2,a_2}) 
\nonumber \\ 	
&=& 
\frac{1}{\sqrt{3}}(\ket{{a_0}',{a_0}'}+ 
\ket{{a_1}',{a_2}'}+\ket{{a_2}',{a_1}'}) 
\label{eq:maxent}
\end{eqnarray}
here the maximally entangled state is described in two different basis, 
the $A$ basis which is the computational basis and the $A'$ basis which 
corresponds to the Fourier transformed:
\begin{eqnarray}
\ket{{a}_{l}'}=\frac{1}{\sqrt{3}}\sum^{2}_{k=0}\exp\left(\frac{2\pi 
i~kl}{3}\right)\ket{{a}_{k}}.
\end{eqnarray}
Notice that the two bases are mutually unbiased, i.e. 
$\braket{{a}_{k}}{{a}_{l}'}=\frac{\exp\left(\frac{2\pi
i~kl}{3}\right)}{\sqrt{3}}$, for 
all $k,l=0,1,2$. 

It is now possible to setup a quantum state targeting game similar to the 
one described in the previous section, but with the twist that Alice and 
Bob share maximally entangled states. One round of the game corresponds to 
using one of the maximally entangled state, hence Alice and Bob  can 
play many times since they share many maximally entangled states. 

In each round of the game i.e. for each qutrit pair, Alice and Bob first 
decide on the set of the two target states (corresponding to 
$\ket{{\psi}_{1}}$ and
$\ket{{\psi}_{2}}$).
 The target states are drawn from two sets of states 
corresponding to the basis states of the two bases $A$ and $A'$; They 
always selects a state from each basis, so that the set of target states 
corresponds to two non-orthogonal states. Notice that the two states 
$\ket{{a}_{k}}$ and $\ket{{a}_{l}'}$ are chosen at random and with equal 
probability and that there in total can be formed nine different sets of 
target states.

\subsection{Alice's measurement}
We are now only at the beginning for the game of quantum state targeting, 
namely where the set of target states are known and Alice has to prepare a 
quantum system to submit to Bob. Here the difference is that Alice has to 
prepare the system by performing a measurement on one part of a maximally 
entangled state. This corresponds to step number (1) in the definition of 
state targeting.

\begin{table}
\begin{center}
\begin{tabular}{|l|l|l|l|} \hline
    & ${a}_{0}'$& ${a}_{1}'$& ${a}_{2}'$ \\ \hline
$a_0$ & $m_{00}$ & $m_{01}$ &$m_{02}$ \\ \hline
$a_1$ & $m_{10}$ & $m_{11}$ &$m_{12}$ \\ \hline
$a_2$ & $m_{20}$ & $m_{21}$ &$m_{22}$ \\ \hline
\end{tabular}
\caption{The intermediate states formed by the basis states from $A$ and 
$A'$}
\end{center}
\end{table}
If Alice want to optimize her control (this means that she is going for 
outcome (A)) we already know that she has to 
perform a measurement such that Bob's qutrit is left in the state which 
corresponds to the intermediate state between the two target states. In 
 table 1 the possible intermediate states are shown. The state  $m_{kl}$ 
corresponds to the intermediate state between
$\ket{{a_k}}$ and  $\ket{{a}_{l}'}$, and with the first index always 
referring
to
the
$A$ and the second to the $A'$-basis, and the general expression for the 
intermediate state $\ket{m_{kl}}$ is:
\begin{eqnarray}
\ket{m_{kl}}=\frac{1}{\sqrt{{\cal 
N}}}\left[\ket{{a_k}}+\exp(\frac{-2\pi i~kl}{3})\ket{{a}_{l}'} \right]
\end{eqnarray}
where ${\cal N}=2(1+1/{\sqrt{3}})$. It should be mentioned that the 
intermediate states in general not are orthogonal.

A measurement on one part of an entangled pair can be considered a 
sophisticated way of making state preparation.

The question is what kind of measurement Alice has to perform on her 
qutrit in order to {\it try} to prepare Bob's qutrit in the desired 
state. It turns out that she has to perform a binary measurement 
corresponding to a measurement of the intermediate states, related to 
the set of target states.

For example if the set of target states is 
$\ket{{a}_{0}}$ and $\ket{{a}_{0}'}$, we know that Alice wants 
to prepare Bob's qutrit in the intermediate state $\ket{m_{00}}$. 
In order to do this Alice 
performs the binary 
measurement 
corresponding to a measurement of  $\ket{m_{00}}$ on her own qutrit. A 
binary measurement is a 'yes'/'no' test, and in this case the question 
asked by Alice is 'is the state of my qutrit $\ket{m_{00}}$?'. The 
question is asked by measuring $\PP_{00}=\ket{m_{00}}\bra{m_{00}}$ 
and 
${\PP}^{\neg}_{00}=\openone - {\PP}_{00}$.  Each of 
the two possible outcomes appears with a certain probability, but the 
important thing is that when the answer to her test is 'yes', i.e. she 
finds the outcome ${\PP}_{00}$,  then she 
knows that she has managed to prepare Bob's qutrit in the desired state. 
Assuming that Alice 
measures on the first qutrit, the state of the system after the measurement, 
is 
(omitting normalization, when not needed), 
\begin{eqnarray}
{\PP}_{00}\ket{\psi_t}&=& 
\ket{m_{00}}\bra{m_{00}} 
\frac{1}{\sqrt{3}} 
\left( \ket{a_0,a_0}+\ket{a_1,a_1}+\ket{a_2,a_2}\right) \nonumber\\
&\propto& \ket{m_{00}}\left( \bra{a_{0}}+\bra{a_{0}'} \right)
\left( \ket{a_0,a_0}+\ket{a_1,a_1}+\ket{a_2,a_2} \right) \nonumber\\
&\propto&\ket{m_{00}}\left( \ket{a_{0}}+\frac{1}{\sqrt{3}}\ket{a_{0}} 
+\frac{1}{\sqrt{3}}\ket{a_{1}}+ 
\frac{1}{\sqrt{3}}\ket{a_{2}}\right)\nonumber\\
&\propto&\ket{m_{00}}\left( 
\ket{a_{0}}+\ket{a_{0}'}\right) \propto\ket{m_{00}}\ket{m_{00}}
\end{eqnarray}
Similarly 
can be shown for the other choices of sets of target states. 

However,
notice the form of the maximally entangled state when written in the two 
bases $A$ and $A'$.  In the $A$ basis there is complete agreement between 
the states obtained by Alice and Bob if they both measured this basis, for 
example if Alice finds the state $\ket{{a_1}}$, Bob too will find the 
state $\ket{{a_1}}$. Whereas if they both measured the $A'$ basis and 
Alice obtains  $\ket{{a}_{1}'}$, then Bob will obtain $\ket{{a}_{2}'}$. 
It is important to understand that this does not give rise to any 
problems, since the important point is that Alice and Bob in both bases 
are perfectly correlated. It only means that Alice has to keep this in 
mind when she choose her measurement; for example when Bob chooses the 
set of target states to $\ket{{a_1}}$ and $\ket{{a}_{2}'}$, Alice knows 
that for her $\ket{{a}_{2}'}$ corresponds to $\ket{{a}_{1}'}$ due to the 
way they are correlated in the $A'$ basis.

\subsection{Bob's choice of target state, Alice's announcement and the 
final test}
At this point in the game Alice has performed a measurement on her qutrit 
in order to try to prepare Bob's qutrit  in  the intermediates state 
of the 
two target states, in this way Alice is optimizing her 
control. This corresponds to step (1), but with Alice determined to go 
for outcome (A). 

Now Bob announces to Alice which of the two states is the target 
state of this round of the game, which means that Bob announce either 
$\ket{{a}_{k}}$ or $\ket{{a}_{l}'}$, this corresponds to step (2) where 
Alice learns the identity of the target state. 
Step (3) is an announcement from Alice and depending on the outcome of her 
measurement Alice will do one of two things: If she succeeded in her 
preparation  so that Bob holds the state $\ket{{m}_{kl}}$, 
then she will 
announce the same state as Bob, i.e. the target state, whereas if she 
has 
failed in making the desired preparation, then she will 
decline
in 
making 
an announcement. Notice that Alice's decline option is completely 
independent of Bob's choise of target state, actually as soon as Alice has 
performed her 
measurement she knows if she will make an announcement or if she will 
decline\footnote{In \cite{RS} there is also decribed a situation where 
Alice has the option to decline, but in that case it is dependent on 
Bob's choise of target state.}.

The cases of interest are of course when Alice has succeeded in making the 
right preparation and makes an announcement. Since we are considering 
the 
case where Alice is going for optimal control and hence is going for 
outcome (A), it means that she is always announcing the target state. 
This means that  
if Bob has chosen $\ket{{a}_{k}}$, Alice will announce that she prepared 
the state $\ket{{a}_{k}}$, whereas if 
Bob has chosen the state  $\ket{{a}_{l}'}$, Alice will claim that she 
prepared the state $\ket{{a}_{l}'}$. 

The final step (step (4)) in the game of quantum state targeting is for 
Bob to make a test of Alice's announcement regarding the state. One way 
that he can make the test is by measuring either $A$ or $A'$ depending 
on 
the chosen target state. Since Alice in this scenario always announces 
the 
target state, it means that if the  target state was $\ket{{a}_{k}}$ then 
Bob will measure $A$. Naturally since the state of Bob's qutrit is not 
$\ket{{a}_{k}}$, but $\ket{{m}_{kl}}$, there is a certain probability that 
Bob will actually find state $\ket{{a}_{k}}$ and hence confirm Alice's 
announcement, but there is also a certain probability that Bob will obtain 
one of the other basis states for $A$, which means that Alice has failed 
Bob's test. Similar for the any other target state.

The probability that Alice's 
announcement will pass Bob's 
test is $p(s)=\frac{1}{2}+\frac{1}{2\sqrt{3}}$, whereas the total 
probability that her announcement fails his test is 
$p(f)=\frac{1}{2}-\frac{1}{2\sqrt{3}}$. Just as in the version of 
quantum state targeting using single particles, the only difference is 
that Alice has the option to 
decline making a statement. 

\section{From quantum state targeting to a Bell inequality for qutrits}
\subsection{The quantum mechanical limit}
In the previous section it was shown how quantum state targeting can be 
played with the use of maximally entangled states, when considering 
the particular situation where Alice is going for maximum control but 
at the same time is given the option to decline to make an announcement. 

The measurements which 
would give the maximum control to Alice was found: she performs a binary 
measurement given by the intermediate state corresponding to  
 the set of target states. Bob for his part is just performing a 
standard von Neumann measurement corresponding to one of the two mutually 
unbiased bases $A$ or $A'$. 

As already mentioned there exist nine different sets of target states
and to each set of target states corresponds one particular binary
measurement for Alice, namely a measurement of the intermediate state
between the two possible target states. Alice can either succeed or fail
in achieving what she wants by her measurement. From hereon we only
consider the cases where Alice succeeds and makes an announcement, in
other words when she managed to perform the desired 'preparation' of
Bob's state.

Bob, on the other hand, has two possible measurements $A$ and $A'$, 
which one he performs depends on his choice of target state. 
In either case there are 
three possible outcomes of the measurement; with a certain probability 
Bob  
will find the target state and confirm Alice's announcement, but there is 
also a certain probability that he will find one of the other two basis 
states and hence fail to confirm Alice's announcement. In total this gives 
$9 \times 2 \times 3 =54$ different probabilities. 

Suppose that Alice and Bob share many maximally entangled states and 
they 
play the game of quantum state targeting many times, then we can 
 ask the question: 'How well is Alice 
playing this 
game?' Here 
we will  measure how well Alice is doing by summing all the 
probabilities when Bob confirms Alice's announcement and from this sum 
subtract all the probabilities when Alice's announcement fails Bob's 
test, in other words
\begin{eqnarray}
B_3 &=& \sum p(Alice ~passes ~Bob's ~test) -\sum p(Alice ~fails ~Bob's 
~test) \nonumber \\
&=&\sum p(s) -\sum p(f) \nonumber
\end{eqnarray}
It is by ordering 
all the involved  probabilities in this particular way that we can 
obtain a Bell 
inequality for two entangled qutrits. In order to write down the Bell 
inequality it is, however, convenient to assign values to the various 
states and 
organize them in different sets, since this simplifies the 
notation. 

\begin{table}
\begin{center}
\begin{tabular}{|l|l|l|l|l|l|} \hline
value & $A$ & $A'$ & $M_0$ & $M_1$ &$M_2$ \\ \hline
0 & $\ket{a_0}$ &$\ket{{a}_{0}'}$ &$\ket{m_{00}}$& $\ket{m_{01}}$&
$\ket{m_{02}}$ \\\hline
1 & $\ket{a_1}$ &$\ket{{a}_{1}'}$ &$\ket{m_{11}}$& $\ket{m_{12}}$&
$\ket{m_{10}}$ \\\hline
2 & $\ket{a_2}$ &$\ket{{a}_{2}'}$ &$\ket{m_{22}}$& $\ket{m_{20}}$&
$\ket{m_{21}}$ \\\hline
\end{tabular}
\caption{Values assigned to the involved states}
\end{center}  
\end{table}

In table 2, it is shown the value which is assign to the 
various states. Notice that the value
for the $A$ and $A'$ basis states  is given by their index. 
The intermediate states has been 
organized into 
three different sets so that the value of the states is always 
given by the first index (i.e. corresponding to the value of the state 
from the $A$ basis). 
The intermediate states are not orthogonal and the sets $M_0$, $M_1$ and 
$M_2$ do therefore {\bf not} correspond to orthogonal bases; the states 
are only 
organized in sets in order to simplify the notation.

In the following we only consider the cases where Alice obtains a useful 
result from her measurement and hence makes an announcement. Suppose 
that the set of target states was chosen such that Alice has measured 
one of 
the projectors in the first set, $M_0$, and Bob following makes a 
measurement in the $A$ basis, for this combination of measurements, 
there are the following contributions to the sum $B_3$:
\begin{eqnarray}
P(A=M_0) &=&\sum_{i=0}^{2}p(a_i 
\cap m_{ii})=\frac{1}{2}+\frac{1}{2\sqrt{3}}\\
P(A\neq M_0)&=&\sum_{i,j=0,j\neq 
i}^{2}p(a_j \cap m_{ii})=\frac{1}{2}-\frac{1}{2\sqrt{3}}
\end{eqnarray}
where $P(A=M_0)$ should be read as follows: after Alice's announcement, 
Bob measures $A$ and finds the correct state and hence he can confirm 
Alice's statement. Contrary $P(A\neq M_0)$ means that Bob finds the 
wrong state and  fails to confirm Alice's announcement. The 
probability $p(a_i \cap m_{ii})$ is the joint probability for obtaining 
both $m_{ii}$ and $a_i$. 

Similar for any of the other measurements that Alice can perform as long 
as Bob always chooses to measure in the $A$ basis after Alice's 
announcement. The same is also the case if the choice of target states 
is such that Alice will measure one of the projectors from the $M_0$ set 
and Bob subsequently measures in the $A'$ basis. In total this leads to 
the following contributions: $P(A=M_i) 
=P(A'=M_0) =\frac{1}{2}+\frac{1}{2\sqrt{3}}$, and $P(A\neq M_i)
=P(A'\neq M_0) =\frac{1}{2}-\frac{1}{2\sqrt{3}}$.

The situation changes when Alice has measured one of the 
projectors from $M_1$ or $M_2$ and Bob subsequently measures $A'$. This 
is due to the fact that in the $A'$ basis Alice and Bob are correlated 
in a different way, as can be seen in eq.(\ref{eq:maxent}). However this 
point has already been discussed (see Sec. 3.1) and does not give rise 
to 
any problems --- it just has to be taken into account. 
Consider the case where Alice has measured one of the projectors from 
$M_1$ and Bob measured in $A'$, then the state that Alice possesses will 
consistently have a value which is $2$ (mod 3) higher than the value of 
Bob's 
state. 
To see this, assume for example that the set of target states were 
$\ket{{a}_{2}}$ and $\ket{{a}_{0}'}$, this means that in the case that 
Alice has succeeded in her projection, she has the state 
$\ket{{m}_{20}}$ which has been assigned the value $2$. At the same 
time, if Bob has chosen target state $\ket{{a}_{0}'}$ and his qutrit 
passes the test for being in that state, Bob will find a state which has 
been assigned the value $0$. Similarly for the other combinations, 
which leads to $P(M_1=A'+2)=\frac{1}{2}+\frac{1}{2\sqrt{3}}$ and 
$P(M_1\neq A'+2)=\frac{1}{2}-\frac{1}{2\sqrt{3}}$. Whereas in the 
situation that Alice uses one of the projectors from $M_2$ and Bob 
subsequently measures in $A'$, Alice will consequently find a value 
which 
is $1$ higher than the value which correlates her with Bob, i.e. 
$P(M_2=A'+1)=\frac{1}{2}+\frac{1}{2\sqrt{3}}$ and
$P(M_2\neq A'+1)=\frac{1}{2}-\frac{1}{2\sqrt{3}}$. 
 
Now all the probabilities can be written down in a nice and compact way, 
and it is easy to evaluate the total sum, based on the the specific set 
of measurements used above:
\begin{eqnarray}
B_3&=&\sum_{i=0}^{2}P(M_i=A)-\sum_{i=0}^{2}P(M_i\neq A)\nonumber\\
   && +\sum_{i=0}^{2}P\left( M_i=A'+\left( 
   3-i\right) \right)-\sum_{i=0}^{2}P\left( M_i\neq 
   A'+\left(3-i\right) \right)\nonumber\\
   &=& 
   2\times 
   3\left(\left(\frac{1}{2}+\frac{1}{2\sqrt{3}}\right)+
   \left(\frac{1}{2}-\frac{1}{2\sqrt{3}}\right)\right)\nonumber \\
   &=&2\sqrt{3}
\end{eqnarray}

\subsection{The local variable limit}
In order to show that the sum of probabilities $B_3$, is actually a Bell 
inequality it is necessary to consider what happens when a local 
variable model tries to attribute definite values to the observables use 
in the game. 

Alice performs binary measurements where each ${m}_{kl}$ is measured 
independently, and hence in a local variable model they could all be 
true simultaneously. Bob on the other hand performs standard von Neumann 
measurements where ${a}_{0}$, ${a}_{1}$ and ${a}_{2}$ are measured 
simultaneously in a single measurement $A$, which means that only one 
of them can come out true in a local variable model. The same is the 
case for ${a}_{0}'$, ${a}_{1}'$ and ${a}_{2}'$ which are measured
simultaneously in a single measurement $A'$. This means that if $a_i$ 
is true, meaning a measurement of $A$ will result in the outcome $a_i$, 
then all probabilities involving $a_j$ with $j\neq i$ must be zero. 

Suppose that in a specific local variable model $a_i$ and ${a}_{j}'$ are 
true, in principle all the $m_{kl}$ can be true at the same time. What 
we need to investigate is what are the contributions from the various 
$m$-states. First notice that the only $m$-state which gives a positive 
contribution to $B_3$ is the one which identifies both $a_i$ and 
${a}_{j}'$ correctly, in other words $m_{ij}$. This will give a 
contribution of $+2$. Whereas $m_{kj}$ and $m_{il}$, where only one 
index is correct will identify only one state correctly and the 
other one wrong; which in total will give a zero contribution to $B_3$. 
Instead the case where both indexes are wrong, and hence both states 
will be wrongly identified will give a negative contribution of $-2$ to  
$B_3$. 

This means that the total maximum of the sum $B_3$ according to a local 
variable model is 
\begin{eqnarray}
B_3 \leq 2
\end{eqnarray}

\section{Quantum state targeting leads to maximal violation}
In the previous section we saw that an attempt to try to assign definite 
values to the observables will lead to $B_3 \leq 2$. But we have also 
seen (Sec. 4.1) that with the measurement settings proposed above and 
using the 
maximally entangled state, it is possible quantum mechanically to obtain 
 $2\sqrt{3}$, which is obviously higher than $2$. 
This means that we have obtained a Bell inequality and moreover shown 
that it can be violated with a $\sqrt{3}$. 

However some important issues still need to be addressed; {\it What is 
the 
maximal violation? For which state is it maximally violated?} and {\it 
What are the optimal measurement settings for the above inequality?} 
These questions have received numerical attention 
\cite{mybell1,mybell2}; first of all 
it has 
been checked that $2\sqrt{3}$ indeed is the quantum mechanical limit to 
the above sum of probabilities and that this maximum is reached for the 
maximally entangled state. Moreover, it has also been checked using 
"polytope software" \cite{poly} that the inequality $B_3\leq 2$ is 
optimal 
for the 
measurement settings presented. 
 
We will try to give some intuitive arguments why this
inequality possesses these properties. The inequality $B_3$, was 
originally
developed to mimic as closely as possible the well-known
Clauser-Horne-Shimony-Holt (CHSH) inequality \cite{CHSH} for qubits. 
This 
means that
there were some very deliberate choices regarding for example the
structure. Very often the CHSH inequality is presented in terms of the
correlation coefficients, which basically is the sum of the
probabilities for the results of the measurements being correctly
correlated while subtracting the probabilities that the results are not
correctly correlated. Hence the choice of the structure of the 
inequality can be considered a generalization of the CHSH inequality. 
And indeed by playing the game of quantum state targeting with qubits 
as it  has been played above with qutrits, will indeed lead to the CHSH 
inequality.

Now the question is why the above inequality is actually optimal for the 
maximally entangled state and this odd choice of measurements, with two 
mutually unbiased basis on one side and nine binary measurements on the 
other! The answer at least intuitively lies in the interplay 
between this state and these measurement settings. 
The choice of two mutually unbiased bases as the possible measurement 
setting on one side means that there is no privileged state. Furthermore 
this symmetry is also preserved by the maximally entangled state. 
Indeed, there exists another Bell inequality for two entangled qutrits 
\cite{otherbell4}, which breaks the symmetry in the choice of bases and  
the 
maximal violation is reached for a pure but non-maximally entangled 
state.

Having chosen the measurement settings on one side, i.e. the two 
mutually unbiased bases $A$ and $A'$ and the maximally entangled state, 
we need to analyze what is the situation on the other side. It has 
already been discussed that we are trying to build a Bell inequality 
with a particular structure, which we can write as
\begin{eqnarray}
B_3 = \sum p(results~correlated) -\sum p(results~not
~correlated) \nonumber
\end{eqnarray}
It is therefore clear that measurement setting on the other side has to 
correspond to the measurements which optimize the probability for 
Alice's 
and Bob's measurement results to be correlated --- independently of 
whether the basis $A$ or $A'$ is being measured. Measuring the 
intermediate states indeed optimize the probability of being correlated, 
or correctly identifying the state as was discussed in section 2 about 
quantum states targeting.

\begin{figure}[t]
\begin{center}
\begin{picture}(85,85)(7.5,7.5)
\put(50,50){\vector(0,1){30}}
\put(50,50){\vector(0,-1){30}}
\put(50,50){\vector(1,0){30}}
\put(50,50){\vector(-1,0){30}}
\put(50,50){\vector(1,1){21.21}}
\put(50,50){\vector(1,-1){21.21}}
\put(50,50){\vector(-1,1){21.21}}
\put(50,50){\vector(-1,-1){21.21}}
\put(82,50){$a_0$}
\put(16,50){$a_1$}
\put(50,82){$a_{0}'$}
\put(50,16){$a_{1}'$}
\put(73,71){$m_{00}$}
\put(73,29){$m_{01}$}
\put(21,71){$m_{10}$}
\put(21,29){$m_{11}$}
\end{picture}
\end{center}
\caption{The optimal measurement settings for the CHSH-inequality in 
two dimensions, 
drawn on the Block-sphere, but labeled according to the notation used 
for intermediate states. Remember that vectors pointing in opposite 
direction are orthogonal.}
\end{figure}
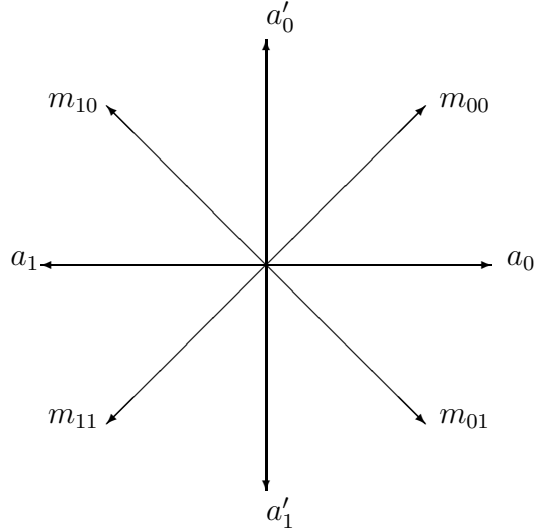

One very interesting point should be mentioned. In two dimensions, which 
means for qubits, the intermediate states actually turns out to be 
pairwise orthogonal and they therefore constitutes two bases. Even more 
the two intermediate bases are also mutually unbiased. This means that 
in two dimensions the Bell inequality which is obtained by performing 
the game of quantum state targeting, corresponds exactly to the 
CHSH-inequality with the optimal measurement settings.  

In three and more dimensions, the situation becomes much more 
complicated, this is due to the fact that the intermediate states in 
general are not orthogonal. However, this problem is overcome by 
considering the intermediate states as binary measurements. But notice 
one interesting point, {\it if} the intermediate states {\it would have} 
formed orthogonal bases, then in three dimensions we would have had that 
when 
generalizing the CHSH inequality in this way, it would have resulted in 
two measurement settings on one side and {\it three} on the other! 

\section{Conclusions}
In recent years, with the field of quantum information rapidly 
expanding, many aspects of quantum mechanics have received renewed 
interest. Here we have considered two things which a priory seem 
unrelated,  namely the game of quantum state targeting and Bell 
inequalities. 

Quantum state targeting is a particular way of combining quantum state 
estimation and additional classical information to obtain an interesting 
quantum game. Here we have taken the starting point that the players 
share many maximally entangled states of two qutrits, and that the 
two possible target states in each round is drawn at random from two 
mutually unbiased bases, $A$ and $A'$. Alice in this case 'submits' her 
quantum system by performing a measurement on her part of the maximally 
entangled state, since this can be viewed as state preparation for Bob's 
part. 

In the case that Alice goes for maximum control, her measurement 
corresponds to measuring the intermediate state between the target 
states and following Bob's announcement of his choice of target 
state she announces the same state. This means that Bob will measure 
either $A$ or $A'$, depending on whether he has chosen the target state 
from $A$ or $A'$.  Alice's measurement of the intermediate state 
corresponds to a binary measurement, in the case where Alice obtains a 
positive result of her measurement she will continue to play the game, 
whereas in the case where she obtains a negative result she will decline 
to make an announcement of the state. It has been proven that for Alice 
to submit what corresponds to the intermediate state between the target 
states is what gives her the optimal control. 

Following this we considered only the cases where Alice succeeds in the 
correct state preparation. Considering all the different measurement 
combinations, i.e. all possible sets of target states and all possible 
choices of target states, we formed a measure of how well Alice is in 
playing this game, by summing over all the probabilities that she will 
pass Bob's test and from this subtract all the probabilities for 
failing his test. Assuming that Alice and Bob are playing by using the 
maximally entangled state and that Alice is measuring on of the nine 
possible intermediate states in a binary measurement and that Bob 
measures one of the two mutually unbiased bases $A$ or $A'$, we find 
that the sum is equal to $2\sqrt{3}$.

Interestingly it turns out the formed sum of probabilities  actually is 
a Bell inequality, since a local variable model would 
for the involved sum of probabilities have a classical limit of $2$. 
Moreover it turns out that the limit we have obtained as a result of 
using 
quantum state targeting is actually the maximum violation, even more the 
maximum violation is reached for the maximally entangled state and for 
the specific measurement setting used in quantum state targeting.

We have argued that the optimality of the obtained inequality arise 
from 
the interplay between the chosen structure of the inequality, the 
symmetry in using mutually unbiased bases together with the maximally 
entangled 
state and finally the intermediate states which optimize the 
probability of guessing correctly the state independently of the chosen 
basis. However, it should be stressed that  these intuitive arguments 
are 
fully supported by numerical evidence.

Here we have played the game of quantum state targeting in three 
dimensions, and it leads to a Bell inequality for two entangled qutrits. 
However the same game could be played in any dimension $d$, the 
measurement settings in this case would be two mutually unbiased bases 
on one side and $d^2$ binary measurements on the other,  and it would 
again lead to an optimal inequality where the classical limit is $2$, 
but the quantum mechanical limit is $2\sqrt{d}$ \cite{mybell1,mybell2}. 
In other words we have 
described a way of obtaining a Bell inequality in arbitrary dimensions 
which has a violation which increases as $\sqrt{d}$. 
The obtained Bell inequality can be considered a generalization of the 
well-known Clauser-Horne-Shimony-Holt inequality, and indeed playing 
the game of quantum state targeting in two dimensions leads to the CHSH 
inequality with its optimal settings. 

Here we have played a simple quantum game and used its optimal solution 
to arrive at a Bell inequality which in itself is optimal, at least in 
some respects. Notice that 
this is somehow backwards with respect to the usual presentation of Bell 
inequalities. Here we start from the quantum state, the 
measurement settings and a specific structure for the sum of 
probabilities and only afterward show that the classical limit is 
lower than the value which was obtained using quantum mechanics.  
We believe this leads to a whole new way of thinking about Bell 
inequalities 
and that in the future this will prove to be a powerful method to 
construct Bell inequalities.

\section*{Acknowledgments}
On this special occasion I would like to take the opportunity to thank 
Asher Peres for all his suggestions and many interesting 
discussions over the past years.


\end{document}